\documentclass[authoryear]{elsarticle}
\makeatletter
\def\ps@pprintTitle{%
  \let\@oddhead\@empty
  \let\@evenhead\@empty
  \def\@oddfoot{\reset@font\hfil} % Removes the footer with the date
  \let\@evenfoot\@oddfoot}
\makeatother

\usepackage{graphicx} % Required for inserting images
\usepackage{hyperref}

\title{bursty\_dynamics: \\ A Python Package for Exploring the Temporal Properties of Longitudinal Data}

\author[1,2]{Alisha Angdembe}
\author[3]{Wasim A Iqbalf}
\author[4]{Rebeen Ali Hamad}
\author[3]{John Casement}
\author[1,4]{AI-Multiply Consortium}
\author[4,5]{Paolo Missier}
\author[6,7]{Nick Reynolds}
\author[1,2]{Rafael Henkin\corref{cor1}}
\author[1,2]{Michael R Barnes\corref{cor1}}

\cortext[cor1]{Both authors contributed equally to this work.}

\address[1]{Faculty of Medicine and Dentistry, Queen Mary University of London}
\address[2]{Digital Environment Research Institute, Queen Mary University of London}
\address[3]{Bioinformatics Support Unit, Faculty of Medical Sciences, Newcastle University, Newcastle upon Tyne, UK}
\address[4]{School of Computing, Newcastle University, Newcastle upon Tyne, UK}
\address[5]{School of Computer Science, University of Birmingham, Edgbaston, UK}
\address[6]{Institute of Translational and Clinical Medicine, Faculty of Medical Sciences, Newcastle University, Newcastle upon Tyne, UK}
\address[7]{NIHR Biomedical Research Centre and Department of Dermatology, Royal Victoria Infirmary, Newcastle Hospitals NHS Foundation Trust, Newcastle upon Tyne, UK}

\begin{document}

\begin{abstract}
Understanding the temporal properties of longitudinal data is critical for identifying trends, predicting future events, and making informed decisions in any field where temporal data is analysed, including health and epidemiology, finance, geosciences, and social sciences. Traditional time-series analysis techniques often fail to capture the complexity of irregular temporal patterns present in such data. To address this gap, we introduce \texttt{bursty\_dynamics}, a Python package that enables the quantification of bursty dynamics through the calculation of the Burstiness Parameter (BP) and Memory Coefficient (MC).  In temporal data, BP and MC provide insights into the irregularity and temporal dependencies within event sequences, shedding light on complex patterns of disease aetiology, human behaviour, or other information diffusion over time. An event train detection method is also implemented to identify clustered events occurring within a specified time interval, allowing for more focused analysis with reduced noise. With built-in visualisation tools, \texttt{bursty\_dynamics} provides an accessible yet powerful platform for researchers to explore and interpret the temporal dynamics of longitudinal data. This paper outlines the core functionalities of the package, demonstrates its applications in diverse research domains, and discusses the advantages of using BP, MC, and event train detection for enhanced temporal data analysis.

\end{abstract}


\maketitle

\section{Motivation and significance}
 Longitudinal data, characterised by repeated observations of the same subjects over time, is prevalent in various disciplines. Analysing the temporal dynamics of such data can reveal significant patterns and trends. Throughout this paper, we refer to an ‘event’ as a data record that includes a timestamp. Traditional time-series analysis methods often fall short in capturing the complexity of temporal properties in longitudinal data, such as clustering or self-exciting relationships between events that may reflect their underlying aetiology. Therefore, metrics like the burstiness parameter (BP) \citep{Kim2016} and memory coefficient (MC) \citep{Goh_2008} have been developed to address this challenge. The BP quantifies the degree of irregularity in the timing of events, and can be used to quantify periods of rapid, clustered events followed by long gaps of inactivity while the MC measures the temporal dependency between successive events, where future events are dependent on the timing of past events. Together, these metrics offer characterisation of the temporal properties of longitudinal data, and may detect intrinsic features, such as self-exciting events. In many fields, understanding the dynamics of these intrinsic features in time courses can lead to critical discoveries—for example, in epidemiology, bursty patterns in disease outbreaks may suggest environmental or behavioral triggers, while in finance, burstiness may reflect market reactions to global events, or fraud related events. The \texttt{bursty\_dynamics} Python package fills an important gap by providing researchers with the tools to easily compute BP and MC, along with an event train detection method that clusters related events. In some cases it may be more appropriate to measure the burstiness of localised temporally clustered events, rather than an entire time course, so event train detection allows for more focused analysis by isolating periods of meaningful activity, reducing noise and improving the accuracy of insights derived from the data. By enabling detailed quantification and visualisation of bursty dynamics, the package allows researchers to analyse complex temporal data with greater precision, thus offering an accessible yet powerful solution for understanding the irregular temporal properties of longitudinal data.

\newpage

\section{Software description}
\subsection{Software Architecture} 
Figure 1 shows the structure of the package. 

\begin{figure}[ht]
\centering
\includegraphics[width=0.8\textwidth]{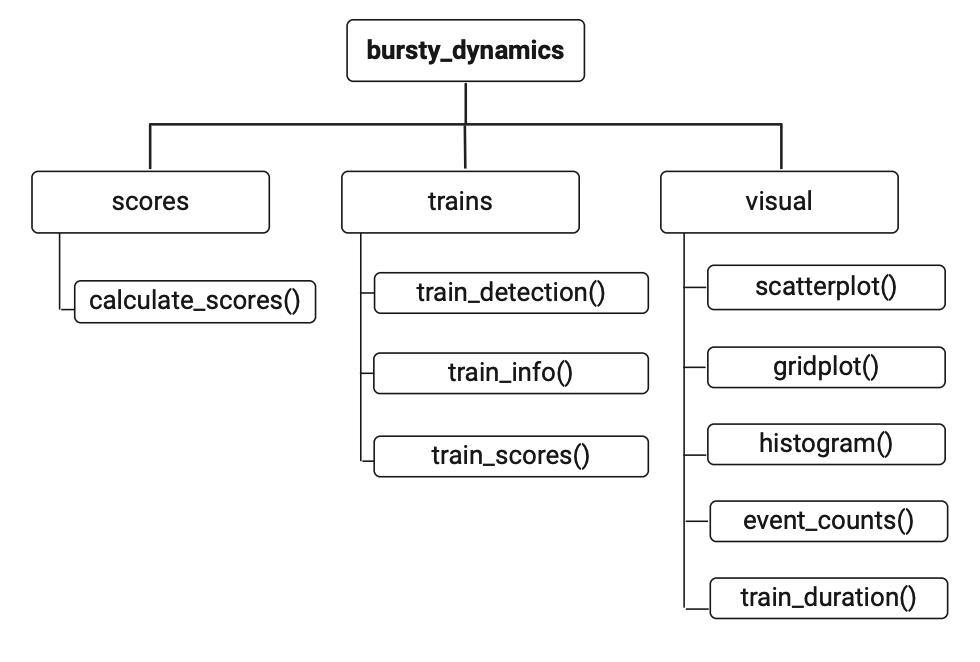} 
\caption{Hierarchical structure of the Python package. The top level represents the package name, followed by the modules \texttt{score}, \texttt{trains}, and \texttt{visual}, with their respective functions listed underneath each module.}
\label{fig1} 
\end{figure}

\subsection{Installation and Usage} 
The \texttt{bursty\_dynamics} package is available on GitHub (\url{https://github.com/ai-multiply/bursty_dynamics}). Users can install the package to a local Python environment. Installation and usage instructions can be found at \url{https://ai-multiply.github.io/bursty_dynamics/}. Once installed, the package provides various functions for analysing temporal properties of longitudinal data directly from the command-line interface or any Python integrated development environment (IDE). Users start by loading their longitudinal data into a DataFrame and using the provided functions to analyse and visualise the temporal patterns.

The input DataFrame must contain the \texttt{subject\_id} and \texttt{time\_col} columns (Table 1). The \texttt{subject\_id} column should include unique identifiers for each subject in the dataset. The \texttt{time\_col} column should contain either date or date and time values in a format that is compatible with Python’s Pandas function \texttt{pandas.to\_datetime}. The package assumes that the input data is clean and pre-processed. Missing values, incorrect data types, and outliers should be handled prior to analysis to ensure accurate results.

\begin{table}[ht]
\centering
\caption{Example of Input Data Format}
\label{tab:input_data}
\begin{tabular}{|c|c|}
\hline
\textbf{subject\_id} & \textbf{time\_col} \\
\hline
1 & 2020-01-01 \\
\hline
1 & 2020-02-01 \\
\hline
2 & 2000-05-01 \\
\hline
\end{tabular}
\end{table}

\subsection{Calculation of Burstiness Parameter and Memory Coefficient} 
The burstiness parameter \citep{Kim2016} is measured using the mean (\(\mu\)) and standard deviation (\(\sigma\)) of the inter-event time distribution, as shown in the equation below:

\begin{equation}
A_n(r) = \frac{\sqrt{n+1} \, r - \sqrt{n-1}}{(\sqrt{n+1} - 2) \, r + \sqrt{n-1}}
\end{equation}

The score ranges between -1 and 1, where a score closer to -1 indicates more regular intervals between events, a score near 0 suggests a random distribution of events, and a score closer to 1 indicates a more severe burst pattern, characterised by rapid, intense occurrences of events as \(\sigma \rightarrow \infty\).

The memory coefficient \citep{Goh_2008} is a statistical measure used to understand the temporal structure of the sequence of events. This coefficient, denoted as \(M\), is calculated using the Pearson correlation formula between successive inter-event times within a given data sequence:

\begin{equation}
M \equiv \frac{\langle T_i T_{i+1} \rangle - \langle T_i \rangle \langle T_{i+1} \rangle}{\sigma_i \sigma_{i+1}}
\end{equation}

\(M\) ranges from -1 to 1, where a positive \(M\) suggests a correlation in the inter-event times (i.e., long times follow long times and short times follow short times), a negative \(M\) implies an alternating pattern between long and short inter-event times, and an \(M\) of 0 indicates no correlation.

Users can calculate BP and MC by passing a DataFrame with subject IDs and event dates to the \texttt{calculate\_scores} function. The package allows optional plotting of scatter plots and histograms for visual analysis. The interpretation of BP and MC is highly domain specific, and users need to have a good understanding of their field to draw meaningful conclusions from these metrics.

\subsection{Event Train Detection} 
Taking inspiration from \citet{CORNER2002127}, we have implemented a method to identify clusters of events, denoted here as \textit{train detection}. We recommend the detection of event trains prior to burstiness analysis in extensive time courses. The \texttt{train\_detection} function detects and assigns train IDs to events based on the specified parameters such as maximum inter-event time and minimum number of events to form a train.

This threshold, the maximum inter-event time, acts as a decisive cut-off point, determining whether two events are close enough in time to be considered part of a series of events with likely shared aetiology. If the interval between two successive events exceeds this threshold, the following event is not included, and the train of events is deemed to have ended.

This flexibility allows for tailored analysis suited to specific research questions. By adjusting parameters and iteratively re-running analyses, users can explore their data, refine their understanding of temporal patterns, and identify meaningful insights in the underlying dynamics of their events. This flexibility allows users to tailor analysis to a specific research question, adjusting parameters to account for domain-specific temporal characteristics. For example, in healthcare data, setting an appropriate inter-event time threshold could help distinguish between symptom flare-ups and unrelated medical events.

\subsection{Analysis of Detected Trains}
The package extends the analysis of the detected trains by calculating BP and MC for each train within each subject's data. The train-level analysis helps to identify patterns within clusters of events, offering deeper insights into the temporal dynamics. The results of the train detection function are sensitive to the chosen parameters (e.g., maximum inter-event time, minimum number of events for a train). Careful tuning and domain-specific knowledge are required to set appropriate values for these parameters.

\subsection{Summary Statistics for Trains}
To provide a comprehensive overview of the detected trains, the package includes a function to calculate summary statistics. This feature aggregates information on train properties, including the number of events per train and the duration of each train. Additionally, it calculates the total number of trains per subject and provides statistical summaries, such as averages and median durations across all trains. 

\newpage

\subsection{Visualisation Tools}
\begin{figure}[ht]
\centering
\includegraphics[width=0.8\textwidth]{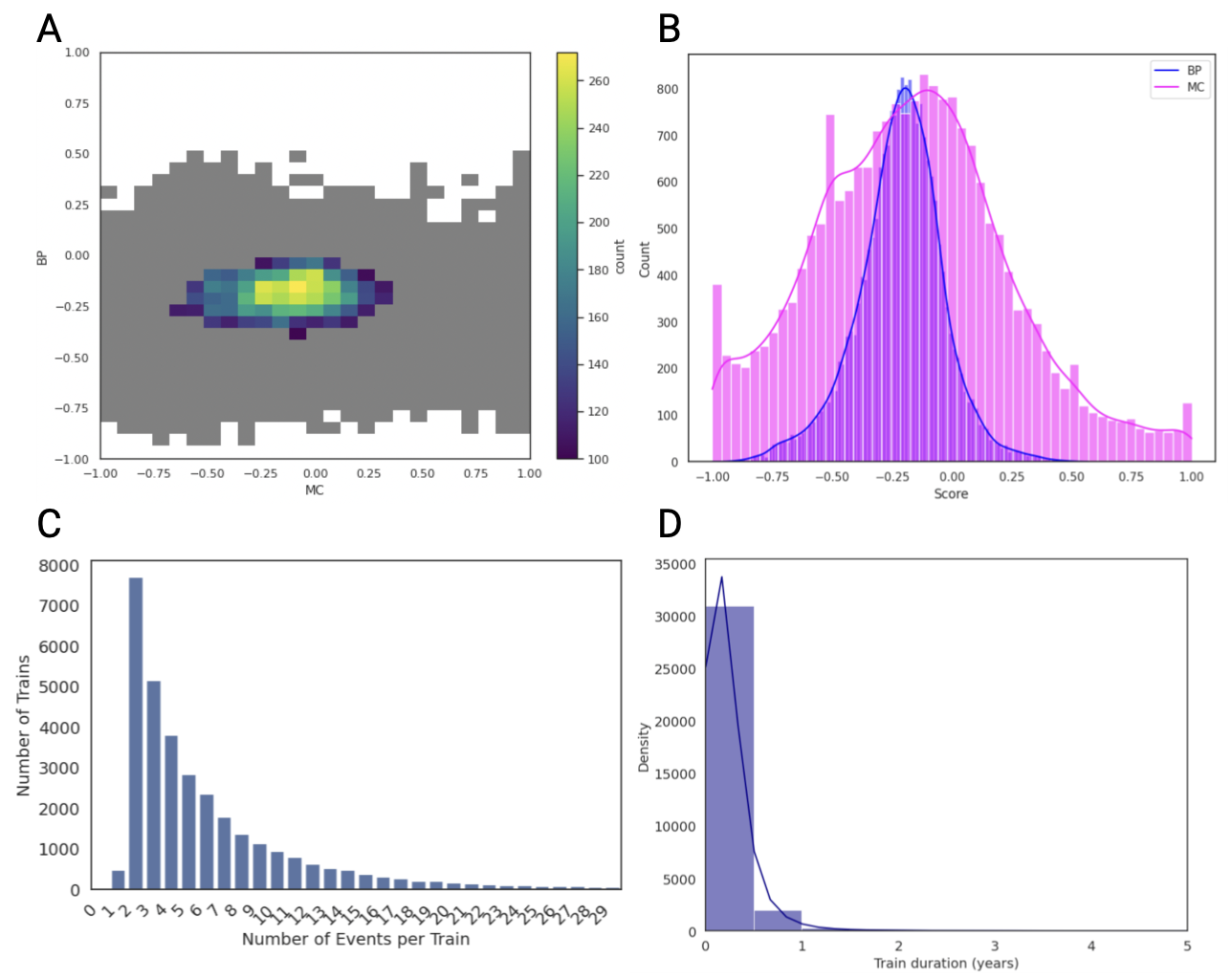}
\caption{\label{fig2}: Examples of visualisations created using the package: (A) Grid plot, (B) Histogram, (C) Event count plot, (D) Train duration plot, using hospital admissions data from MIMIC-IV (\citealp{Johnson2024})}
\end{figure}

The package includes various visualisation tools to aid in the interpretation of the calculated metrics and detected trains:
\begin{itemize}
    \item \textbf{Grid Plot}: Visualise the relationship between MC and BP with a colour bar.
    \item \textbf{Histogram}: Display the distribution of BP and MC, with options of separate or overlapping histograms.
    \item \textbf{Marginal Histograms}: Combine scatter plot with histograms to explore the relationship between MC and BP, and their distributions.
    \item \textbf{Train Duration Plots}: Display the distribution of train duration.
    \item \textbf{Event Count Plots}: Display the count of events per train.
\end{itemize}

The visualisation module offers plotting functions that facilitate group comparisons using the ‘hue’ parameter, enabling the user to colour-encode and segment data based on a specific categorical variable. Functions like \texttt{scatterplot}, \texttt{histogram}, \texttt{train\_duration}, and \texttt{event\_counts} utilise this feature, making it valuable for researchers to uncover insights and present findings clearly. An example of this feature can be seen in the bottom half of the example \href{https://github.com/ai-multiply/bursty_dynamics/blob/main/example/examples.ipynb}{notebook}, under the section \textit{‘Analysing Data with Group Comparisons’}.

While the package provides several visualisation tools, complex visualisations or custom plots may require additional coding and use of external libraries beyond what is provided by the package.

\section{Illustrative Example}
To demonstrate the functionalities of \texttt{bursty\_dynamics}, we use version 3.1 of the MIMIC-IV (Medical Information Mart for Intensive Care) database (\citealp{Johnson2024}), a large repository of time-stamped healthcare data. Additionally, a Jupyter notebook with a working example using a portion of the UK Biobank Synthetic Dataset (\citealp{bio}) is \href{https://github.com/ai-multiply/bursty_dynamics/blob/main/example/examples.ipynb}{available}, including the analysis conducted on event trains.

\subsection{Bursty Dynamics of Emergency Hospital Admissions }
For this exemplar, we used a subset of MIMIC-IV, focusing on patients with at least five hospital admissions to ensure a sufficient sample size for pattern analysis. Patients with fewer admissions were excluded. Using the \texttt{calculate\_scores} function from \texttt{bursty\_dynamics}, we calculated BP and MC for each patient using \texttt{subject\_id} as the identifier and \texttt{admittime} as the timestamp column. The output metrics are visualised through a heatmap generated using the \texttt{gridplot} function. The Python code used for these computations is presented in Figure 3.

The heatmap in Figure 4 provides a visual representation of BP and MC scores for this cohort. Most patients appear in the region where BP $> 0$ and MC $< 0$, indicating that most hospital admissions occur in irregular, bursty patterns, while the intervals between these admissions alternate between long and short periods. It is also important to note that some patients have BP values close to 0, suggesting a random distribution of their hospital admissions over time, while others exhibit MC values near 0, indicating no correlation between their admission intervals, meaning that the timing of their admissions does not follow any consistent pattern.

This example highlights how \texttt{bursty\_dynamics} can distinguish between regular, random, bursty, and alternating temporal patterns in hospital admissions, offering valuable insights into patient healthcare utilisation.

\newpage

\begin{figure}
    \centering
    \includegraphics[width=1\textwidth]{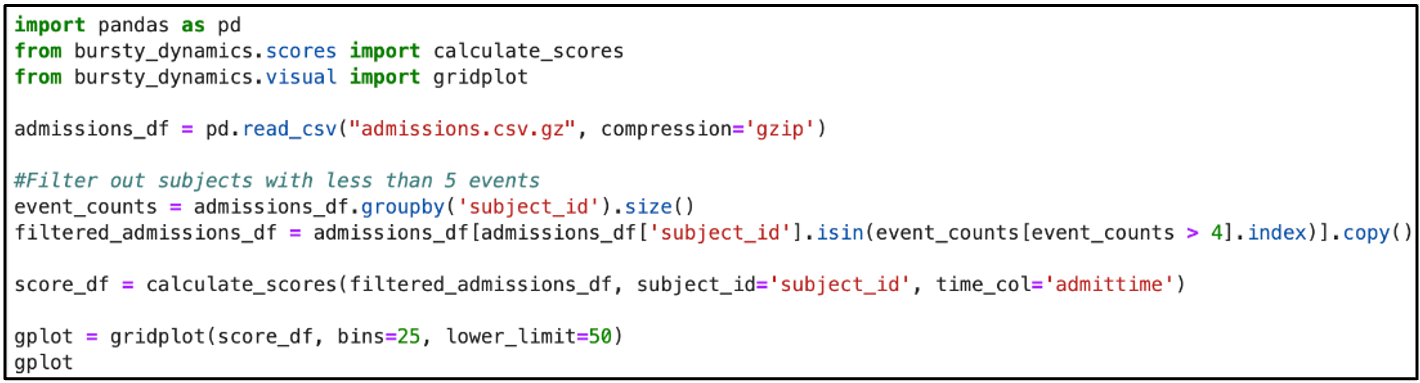} 
    \caption{Python code used for calculating burstiness parameter (BP) and memory coefficient (MC) metrics and generating a heatmap of the result.}
    \label{fig3}
\end{figure}

\newpage

\begin{figure}
    \centering
    \includegraphics[width=1\textwidth]{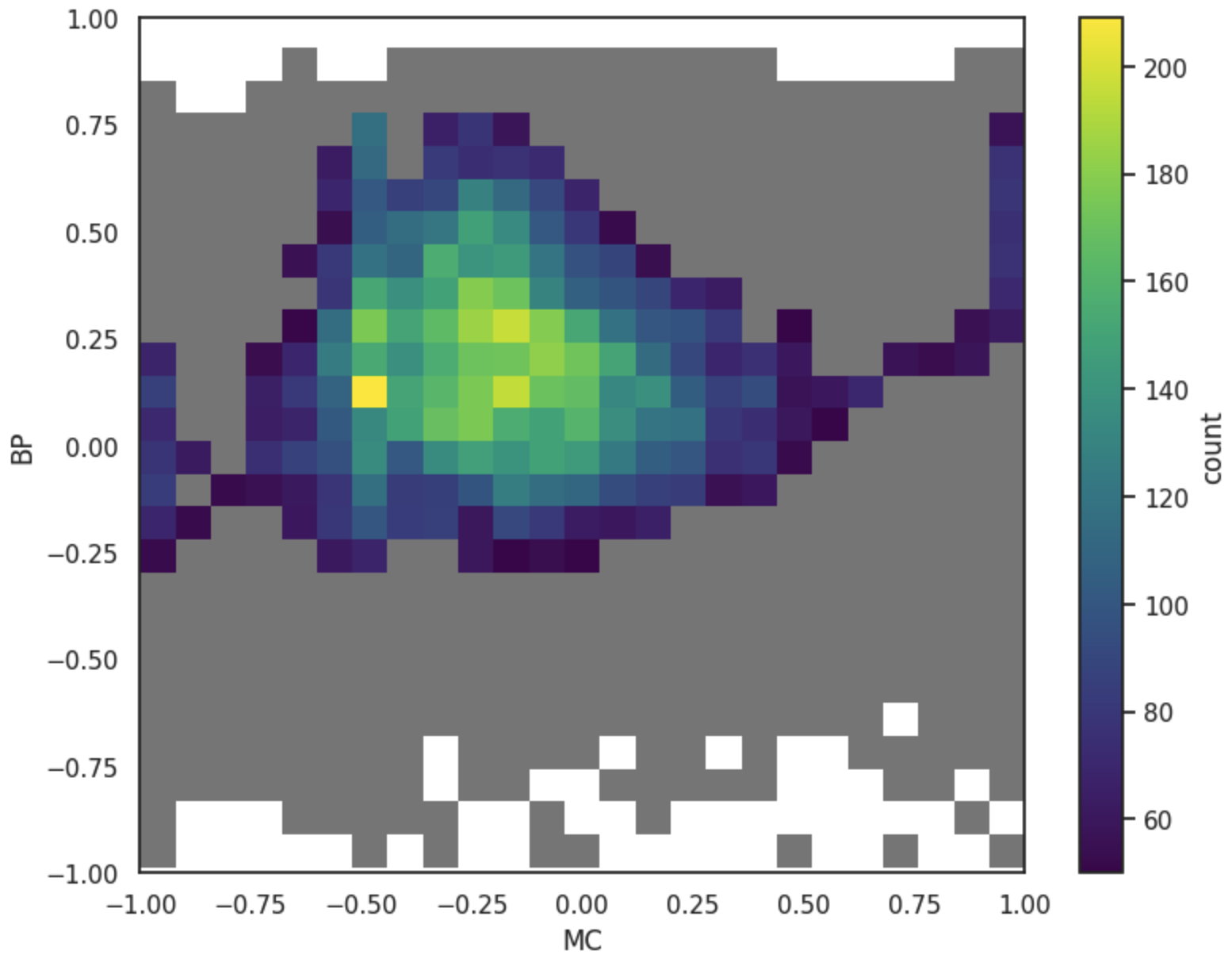} 
    \caption{The heatmap displays the joint distribution of BP and MC for emergency hospital admission events for all subjects, with colour intensity representing the density of the event counts. }
    \label{fig4}
\end{figure}

\clearpage

\subsection{Analysis of Emergency Hospital Admission by Age Group}

For this analysis, we grouped patients based on their age in years at a specific timepoint (in this case 2008) into four age categories: 20-39, 40-59, 60-79, and $\geq$80. We selected 2008 as a reference year because it marks the starting point for patient admissions included in the master patient list, ensuring consistent age calculations across the entire cohort. The BP and MC scores for each age group were calculated based on each patient’s hospital admissions, as in the previous section. The \texttt{histogram} function was utilised to illustrate the distribution of BP and MC scores, using the \texttt{hue} parameter to colour-code by age group, as seen in Figure 5.

The pattern of emergency hospital admission in the figure shows a trend in burstiness that decreases with age. This is most evident in the 20-39 age group, which demonstrates a peak that is shifted slightly to the right of 0, indicating a tendency toward more bursty patterns, while the older age groups are shifted to the left. Across all age groups, the MC distribution exhibits a similar pattern with peaks around -0.5 and 0, especially for the 40-59 and 60-79 age groups. These peaks suggest that hospital admission intervals for these age groups often alternate between long and short periods.

\begin{figure}
    \centering
    \includegraphics[width=\textwidth]{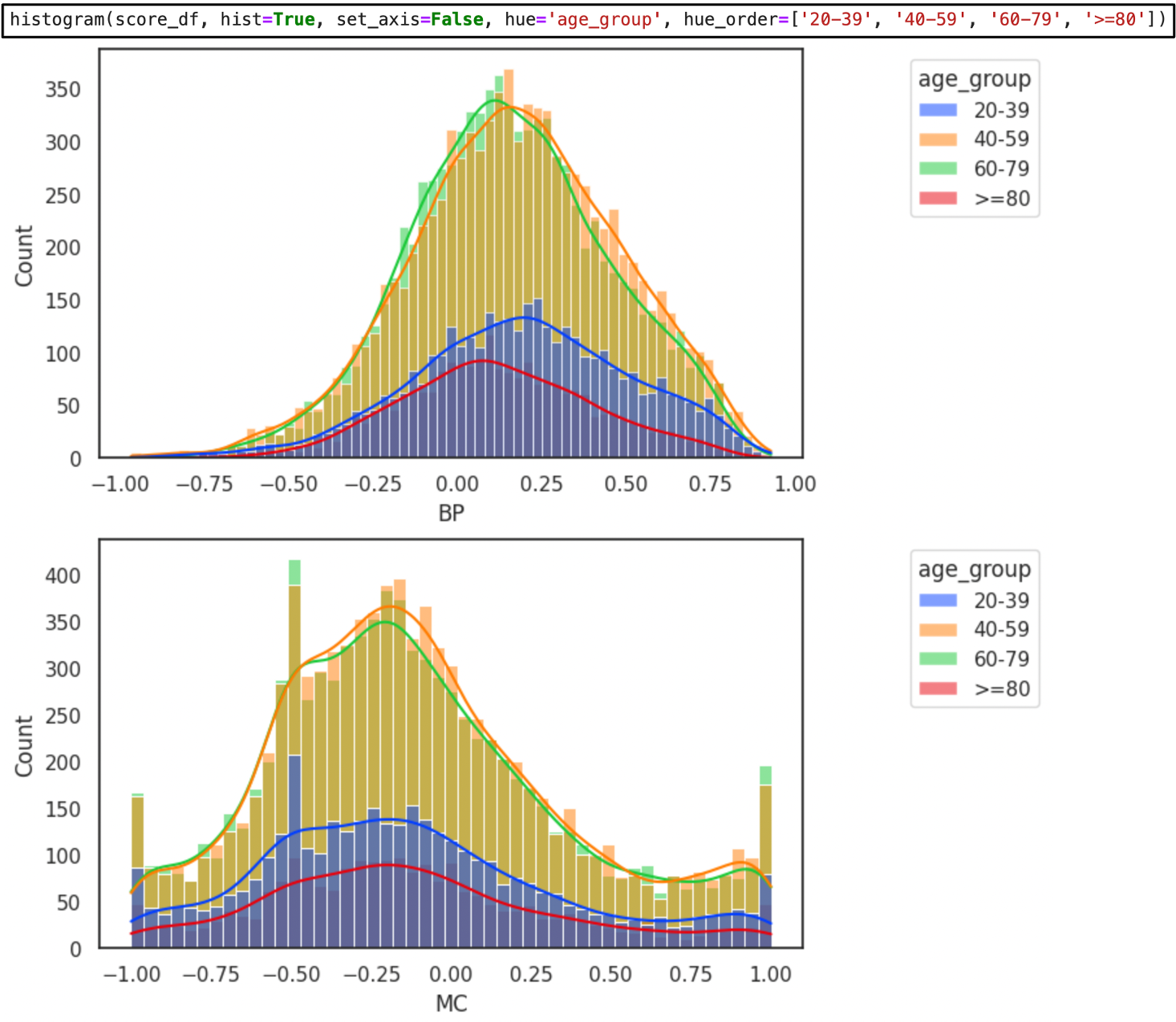} 
    \caption{Python code and histograms visualising the distribution of emergency hospital admission BP and MC by different age groups.}
    \label{fig5}
\end{figure}

\clearpage

\section{Impact}
The \texttt{bursty\_dynamics} Python package addresses a need in temporal data analysis, providing a straightforward, effective solution for the analysis of complex temporal patterns within longitudinal data. By offering tools to calculate the BP and MC, as well as event train detection for grouping temporally related events, the package reveals patterns often missed by traditional methods.

A key feature of \texttt{bursty\_dynamics} is its event train detection, which allows researchers to isolate relevant trains of events specific to their research question. This flexibility enhances insights by identifying periods of high activity, such as symptom flare-ups in clinical data, while minimising noise. Parameter fine-tuning further enables researchers to tailor detection criteria to their field.

In addition to analysis, \texttt{bursty\_dynamics} enhances interpretability with visualisation tools, making it easy for researchers to explore trends visually and compare group patterns. The package’s user-friendly design, detailed documentation, and clear visual summaries make it accessible for users with diverse technical skills.

By providing these tools in a single package, \texttt{bursty\_dynamics} promotes reproducibility, facilitates temporal analysis, and supports domain-specific discoveries, positioning itself as a valuable resource for research.

\section{Conclusion}

The presented Python package offers a user-friendly solution for robust and reproducible analysis of temporal properties of longitudinal data. By quantifying burstiness and memory effects, either in “event trains” or whole-time courses, the package aids researchers in uncovering valuable insights from their data. The visualisation tools further enhance the utility of the package, making it an essential tool for temporal data analysis.

The \texttt{bursty\_dynamics} Python package offers an accessible solution for the analysis of complex temporal patterns in longitudinal data. By quantifying burstiness and memory effects, either in “event trains” or whole-time courses, the package aids researchers in uncovering and visualising valuable insights from their data. The package has great versatility and can be applied to a wide range of research fields where temporal dynamics may play a role.

\section*{Acknowledgements}

This study received funding from the National Institute for Health and Care Research (NIHR), Artificial Intelligence for Multiple Long-Term Conditions (AIM) Development and Collaboration grants, award NIHR203982. The views expressed are those of the author(s) and not necessarily those of the NIHR or the Department of Health and Social Care.

We would like to acknowledge the role of the AI-MULTIPLY Patient and Public Involvement and Engagement (PPIE) group, including Social Action for Health, throughout the project and the discussions leading to the work presented in this paper.

NJR’s research/laboratory is funded in part by the NIHR Newcastle Biomedical Research Centre, the NIHR Newcastle HeathTech Research Centre in Diagnostic and Technology Evaluation and the NIHR Newcastle Patient Safety Research Collaboration. NJR is a NIHR Senior Investigator.

\bibliographystyle{elsarticle-harv}
\bibliography{bib}

\begin{thebibliography}{5}
\expandafter\ifx\csname natexlab\endcsname\relax\def\natexlab#1{#1}\fi
\providecommand{\url}[1]{\texttt{#1}}
\providecommand{\href}[2]{#2}
\providecommand{\path}[1]{#1}
\providecommand{\DOIprefix}{doi:}
\providecommand{\ArXivprefix}{arXiv:}
\providecommand{\URLprefix}{URL: }
\providecommand{\Pubmedprefix}{pmid:}
\providecommand{\doi}[1]{\href{http://dx.doi.org/#1}{\path{#1}}}
\providecommand{\Pubmed}[1]{\href{pmid:#1}{\path{#1}}}
\providecommand{\bibinfo}[2]{#2}
\ifx\xfnm\relax \def\xfnm[#1]{\unskip,\space#1}\fi
%Type = Misc
\bibitem[{Biobank(n.d.)}]{bio}
\bibinfo{author}{Biobank}, \bibinfo{year}{n.d.}
\newblock \bibinfo{title}{Uk biobank synthetic dataset}.
\newblock \bibinfo{howpublished}{\url{https://biobank.ndph.ox.ac.uk/synthetic_dataset/}}.
\newblock \bibinfo{note}{Accessed: 13/05/2024}.
%Type = Article
\bibitem[{Corner et~al.(2002)Corner, {van Pelt}, Wolters, Baker and Nuytinck}]{CORNER2002127}
\bibinfo{author}{Corner, M.}, \bibinfo{author}{{van Pelt}, J.}, \bibinfo{author}{Wolters, P.}, \bibinfo{author}{Baker, R.}, \bibinfo{author}{Nuytinck, R.}, \bibinfo{year}{2002}.
\newblock \bibinfo{title}{Physiological effects of sustained blockade of excitatory synaptic transmission on spontaneously active developing neuronal networks—an inquiry into the reciprocal linkage between intrinsic biorhythms and neuroplasticity in early ontogeny}.
\newblock \bibinfo{journal}{Neuroscience \& Biobehavioral Reviews} \bibinfo{volume}{26}, \bibinfo{pages}{127--185}.
\newblock \URLprefix \url{https://www.sciencedirect.com/science/article/pii/S0149763401000628}, \DOIprefix\doi{https://doi.org/10.1016/S0149-7634(01)00062-8}.
%Type = Article
\bibitem[{Goh and Barabási(2008)}]{Goh_2008}
\bibinfo{author}{Goh, K.I.}, \bibinfo{author}{Barabási, A.L.}, \bibinfo{year}{2008}.
\newblock \bibinfo{title}{Burstiness and memory in complex systems}.
\newblock \bibinfo{journal}{Europhysics Letters} \bibinfo{volume}{81}, \bibinfo{pages}{48002}.
\newblock \URLprefix \url{https://dx.doi.org/10.1209/0295-5075/81/48002}, \DOIprefix\doi{10.1209/0295-5075/81/48002}.
%Type = Misc
\bibitem[{Johnson et~al.(2024)Johnson, Bulgarelli, Pollard, Gow, Moody, Horng, Celi and Mark}]{Johnson2024}
\bibinfo{author}{Johnson, A.}, \bibinfo{author}{Bulgarelli, L.}, \bibinfo{author}{Pollard, T.}, \bibinfo{author}{Gow, B.}, \bibinfo{author}{Moody, B.}, \bibinfo{author}{Horng, S.}, \bibinfo{author}{Celi, L.A.}, \bibinfo{author}{Mark, R.}, \bibinfo{year}{2024}.
\newblock \bibinfo{title}{Mimic-iv (version 3.1)}.
\newblock \bibinfo{howpublished}{PhysioNet}.
\newblock \URLprefix \url{https://doi.org/10.13026/kpb9-mt58}.
%Type = Article
\bibitem[{Kim and Jo(2016)}]{Kim2016}
\bibinfo{author}{Kim, E.K.}, \bibinfo{author}{Jo, H.H.}, \bibinfo{year}{2016}.
\newblock \bibinfo{title}{Measuring burstiness for finite event sequences}.
\newblock \bibinfo{journal}{Phys. Rev. E} \bibinfo{volume}{94}, \bibinfo{pages}{032311}.
\newblock \URLprefix \url{https://link.aps.org/doi/10.1103/PhysRevE.94.032311}, \DOIprefix\doi{10.1103/PhysRevE.94.032311}.

\end{thebibliography}

\end{document}